\documentclass{appolb}
\usepackage{graphicx}
\usepackage{amssymb,amsmath,amsfonts}
\usepackage{latexsym}
\usepackage{graphicx}
\usepackage{caption}
\usepackage{subcaption}
\usepackage{enumerate}
\usepackage{float}
\usepackage{bm}
\usepackage{mathrsfs}
\usepackage{color}
\usepackage{graphics}
\usepackage{epsfig}
\usepackage{bbm}  
\usepackage{soul}   
\usepackage{scalerel,stackengine}
\stackMath
\usepackage{nicefrac}
\usepackage{color}
\usepackage{xcolor}
\usepackage[breaklinks,colorlinks,backref]{hyperref}
\hypersetup{
    colorlinks, 
    linktoc=all, 
    linkcolor=red,  
  citecolor=hyptxt,
  urlcolor=blue}
  \hypersetup{
  citebordercolor=,
  filebordercolor=green,
  linkbordercolor=blue
}
\definecolor{hyptxt}{rgb}{0.7, 0.4, 0.9}

\newcommand\reallywidecheck[1]{%
\savestack{\tmpbox}{\stretchto{%
  \scaleto{%
    \scalerel*[\widthof{\ensuremath{#1}}]{\kern-.6pt\bigwedge\kern-.6pt}%
    {\rule[-\textheight/2]{1ex}{\textheight}}
  }{\textheight}%
}{0.5ex}}%
\stackon[1pt]{#1}{\scalebox{-1}{\tmpbox}}%
}

\newcounter{mnotecount}[section]

\newcommand{\ud}{\mathrm{d}}

\def\ud{\mathrm{d}}

\def\R{\mathbb{R}}

\def\vap{\varpi}

\def\ii{\mathrm{i}}

\def\sfM{\mathsf{M}}
\def\sfMps_{\sfM_{\sigma}^{\vap}}

\def\ud{\mathrm{d}}

\def\bsb{\boldsymbol{\beta}}

\def\ii{\mathrm{i}}

\begin{document}
\title{Mixmaster universe: \\semiclassical dynamics and inflation from bouncing
\thanks{Presented at \textit{The 8th Conference of the Polish Society on Relativity}, Warsaw, Poland, September 19-23, 2022}%
}
\author{Jaime de Cabo Martín
\thanks{jaime.decabomartin@ncbj.gov.pl}
\address{National Centre for Nuclear Research, Pasteura 7, 02-093 Warsaw, Poland}
\\[3mm]
}
\maketitle

\begin{abstract}
    In this work we explore the quantum Bianchi type IX-model, its semi-classical features, and its relevance in early cosmology to tentatively explain inflation and production of primordial structures. We specially focus on the analytical and numerical exploration of the dynamical system derived from the phase-space portraits. Afterwards we investigate the reliability of our results with regard to inflation and post inflation scenarii commonly accepted nowadays.
\end{abstract}

\section{Introduction}
The Mixmaster is an homogeneous model of the early universe, originally studied by Misner \cite{misner69}, in which it underwent an oscillatory and chaotic epoch close to the initial cosmological singularity. 
In the Mixmaster, the spatial slices (positively curved and topologically three-spheres $S^3$) distort anisotropically, since they evolve differently and randomly in each direction. Quantum cosmology combines general relativity with quantum mechanics for the purpose of explaining the
origin of the primordial universe, its expansion and generation of its large-scale structure. In the present work we study a more generic cosmological scenario than the standard approach based on Friedmann cosmology, since from the very beginning we make no assumption  of the approximate isotropy of the primordial space, reducing the number of primordial symmetries. Namely, we drop the isotropy by employing the Bianchi type IX model (corresponding to the ‘Mixmaster universe’). Although not fully generic, the proposed model is significantly less restrictive than the standard one and exhibits a new and complex behaviour on approach to the singularity.

Presently, we devote our full attention to the issue of dynamics of anisotropic cosmological background, regardless of primordial perturbations. We quantize the Bianchi IX model and introduce a semi-classical framework in which its dynamics is much more accessible, though far from trivial. Our quantization procedure respects the symmetries of the phase space of the Bianchi IX model and thereby produces a self-adjoint representation of relevant observables such as the Hamiltonian. The main outcome is the resolution of the big-bang singularity with a bouncing dynamics. The semi-classical framework is derived with the use of coherent states that also respect the mentioned symmetries. 
The semi-classical phase space can be easily showed to exhibit a generic bounce replacing the big-bang and big-crunch singularities.

The work presented in this contribution represents a summary of the forthcoming paper \cite{paper}. In this work we focus on the study of the
role of anisotropy in the semi-classical dynamics close
to the bounce. We try to understand the dynamics by addressing a specific question rather than trying to resolve all
the mathematical difficulties of finding the full solution: ``Can the interplay between anisotropy and semiclassical bounce generate enough accelerated expansion such that it can reproduce a realistic inflationary behaviour that last sufficiently long?" The bouncing isotropic cosmologies produce primordial perturbation amplitude
that generically is blue-tilted contrary to inflationary predictions and the observational results. The existence
of a robust inflationary phase in a bouncing anisotropic
model could therefore provide a serious challenger to the
hypothesis of a primordial scalar field, inflaton, and its
paramount cosmological role \cite{thomasbuchert}. On the other hand, the
non-existence of such a phase in our model, and generally
speaking an inconsequential role of anisotropy, should
strengthen the existing argument in favour of inflaton as
another attempt at disproving its exceptional character
fails.

\section{Classical model}
\label{prelim}
The Hamiltonian formulation of the Bianchi type IX model is described by the following line element \cite{paper}:
\begin{equation}
\ud s^2= -{\cal N}^2\ud\tau^2+\sum_ia_i^2(\omega^i)^2\, ,
\end{equation}
where {$\ud \omega_i=\frac{1}{2} \frak{n}\varepsilon_{i}^{\, jk}\omega_j \wedge \omega_k$, ${\cal N}(\tau)$ and $a_i(\tau)$ are positive-valued functions of time. The Hamiltonian constraint of this spacetime model expressed in the Misner variables $(\Omega,p_{\Omega},\bsb, \mathbf{p})\in\mathbb{R}^6$ reads \cite{misner69}:
\begin{align}\label{con}\begin{split}
&\mathrm{C}=\frac{{\cal N}e^{-3\Omega}}{24}\left(\frac{\mathcal{V}_0}{2\kappa}\right)\left(\left(\frac{2\kappa}{\mathcal{V}_0}\right)^2[-p_{\Omega}^2+\mathbf{ p}^2]+36\frak{n}^2e^{4\Omega}[V(\bsb)-1]\right),\end{split}
\end{align}
where $\bsb:=(\beta_{+},\beta_-)\in \R^2$ and $\mathbf{p}:= (p_+,p_-)\in \R^2$ are canonically conjugate variables, $\mathcal{V}_0=\frac{16\pi^2}{\frak{n}^3}$ is the coordinate volume of the spatial section, $\kappa=8\pi G$ is the gravitational constant. In what follows we set $\frak{n}=1$ and $2\kappa=\mathcal{V}_0$. The gravitational Hamiltonian $\mathrm{C}$ resembles the Hamiltonian of a particle in the 3D Minkowski spacetime moving in a time-dependent potential. The cosmological interpretationof the Misner variables (see \cite{paper} for more details) is such that the variable $\Omega$ describes the isotropic part of geometry, whereas $\beta_{\pm}$ describe the distortions to isotropy and are called the anisotropic variables. The potential that drives the motion of the geometry originates from the spatial curvature, and it reads (Fig. \ref{potentials}-\textit{left}):
\begin{equation}\label{b9pot}
V(\bsb) = \frac{e^{4\beta_+}}{3} \left[\left(2\cosh(2\sqrt{3}\beta_-)-e^{- 6\beta_+}
\right)^2-4\right] +  1 \,.
\end{equation}

Following previous works \cite{berczgamapie15A,berczgamapie15B,berczgama16B,berczgama16A} we redefine the isotropic variables as $q=e^{\frac{3}{2}\Omega}, ~p=\frac{2}{3}e^{-\frac{3}{2}\Omega}p_{\Omega}$. Note that $q>0$ and thus the range of the isotropic canonical variables is the open half-plane that admits the affine group of symmetry transformations, an essential property used in our covariant quantization of the model. The Hamiltonian constraint (\ref{con}) is given by a sum of the isotropic and anisotropic parts, $\mathrm{C}=-\mathrm{C}_{iso}+\mathrm{C}_{ani}$:
\begin{align}\begin{split}
\mathrm{C}_{iso}=\frac{{\cal N}}{24}\left(\frac{9}{4}p^2+36q^{\frac{2}{3}}\right),~~
\mathrm{C}_{ani}=\frac{{\cal N}}{24}\left(\frac{\mathbf{p}^2}{q^2}+36q^{\frac{2}{3}}V(\bsb)\right).\end{split}
\end{align}

\section{Quantum model and its semi-classical portrait}
\label{quantsmod}

\subsection{Covariant Weyl-Heisenberg integral quantization of functions on plane}
We consider a four-dimensional phase space $\R^4=\R^2\times\R^2$ made of two pairs of canonical anisotropic variables, $(\beta_+,p_+)$ and $(\beta_-,p_-)$, and  define the integral quantization of a function $f(\mathbf{r}_{\pm})$ in the phase space $\mathbf{r}_{\pm}=(\beta_{\pm},p_{\pm})\in\mathbb{R}^2$ (we omit the index $_{\pm}$ in the sequel) as the following:
\begin{align}
f(\mathbf{r})\mapsto A_{f}:=\int_{\mathbb{R}^2} f(\mathbf{r})\mathcal{Q}(\mathbf{r})\frac{\ud^2\mathbf{r}}{2\pi},
\end{align}
A non-trivial UIR of the three-dimensional Weyl-Heisenberg group is $
U(\mathbf{r})=e^{\ii(pQ-\beta P)}$,
where $Q$ and $P$ are the usual essentially self-adjoint position and momentum operators on the line with $[Q,P]=\ii \hbar \mathbbm{1}$. It turns out that any admissible family of operators $\mathcal{Q}(\mathbf{r})$ has the form $
\mathcal{Q}(\mathbf{r})=U(\mathbf{r})\mathcal{Q}_0U(\mathbf{r})^{\dagger},$
where $\mathcal{Q}_0$ is a unit-trace operator on $\mathcal{H}$. Thus, the choice of a quantization procedure is reduced to the choice of a single operator, $\mathcal{Q}_0$. Equivalently, one may use the weight function, $\Pi(\mathbf{r})$, which is defined via the Weyl-Heisenberg transform of $\mathcal{Q}_0:~$ $\Pi(\mathbf{r}):=\mathrm{Tr}(U(-\mathbf{r})\mathcal{Q}_0)~~\Longrightarrow ~~\mathcal{Q}_0=\int_{\mathbb{R}^2}U(\mathbf{r})\Pi(\mathbf{r})\frac{\ud^2\mathbf{r}}{2\pi}$
to determine the quantization procedure. We must assume $\Pi(0)=1$. The weight $\Pi(\mathbf{r}_{\pm})$ defines the extent of coarse graining of the phase space $\mathbf{r}_{\pm}=(\beta_{\pm},p_{\pm})\in\mathbb{R}^2$.

\subsection{Semi-classical portraits}
Given a quantum operator $A_f$ corresponding to the observable $f$, we define the so-called quantum phase space portrait of that operator by making use of another family of bounded unit-trace operators $\mathcal{R}(\mathbf{r})$ that we use for quantization. Also, more tractable formulas are obtained when the weight function $\Pi(\mathbf{r})$ instead of the family of operators $\mathcal{Q}(\mathbf{r})$. Thus, we obtain:
\begin{align}
\reallywidecheck{f}(\mathbf{r})=\mathrm{Tr}(\mathcal{R}(\mathbf{r})A_f)=\int_{\mathbb{R}^2} \mathcal{F}[\Pi]*\mathcal{F}[\widetilde{\Pi}](\mathbf{r}'-\mathbf{r})f(\mathbf{r}')\frac{\ud^2\mathbf{r}'}{4\pi^2}.
\end{align}

\subsection{Semi-classical portrait of the anisotropy}
Assuming the Gaussian weight function $\Pi(\beta,p)=e^{-\frac{\beta^2}{\sigma^2}}e^{-\frac{p^2}{\omega^2}}$
where the width parameters $\sigma$ and $\omega$ encode our degree of confidence in dealing with a given point in the phase space, we easily find for the anisotropic momentum and potential (Fig. \ref{potentials}-\textit{right}):
\begin{eqnarray}
&\reallywidecheck{(p^2)}=p^2+\frac{8}{\sigma^2},\\
\reallywidecheck{V}&(\beta_{\pm})=\frac{1}{3}\left(D(4\sqrt{3},4)e^{4\sqrt{3}\beta_-+4\beta_+}+D(4\sqrt{3},4)e^{-4\sqrt{3}\beta_-+4\beta_+}+D(0,8)e^{-8\beta_+}\right)\nonumber\\ &-\frac{2}{3}\left(D(2\sqrt{3},2)e^{-2\sqrt{3}\beta_--2\beta_+}+D(2\sqrt{3},2)e^{2\sqrt{3}\beta_--2\beta_+}+D(0,4)e^{4\beta_+}\right)+1, \label{semipot}
\end{eqnarray}
where the $D(x,y)=e^{\frac{4x^2}{\omega_-^2}}e^{\frac{4y^2}{\omega_+^2}}$ are regularization factors issued from our choice of the Gaussian weights. For a more detailed explanation of the quantization process and derivation of semiclassical portrait see \cite{paper}.

\begin{figure}[h!]\centering
\includegraphics[width=0.8\textwidth]{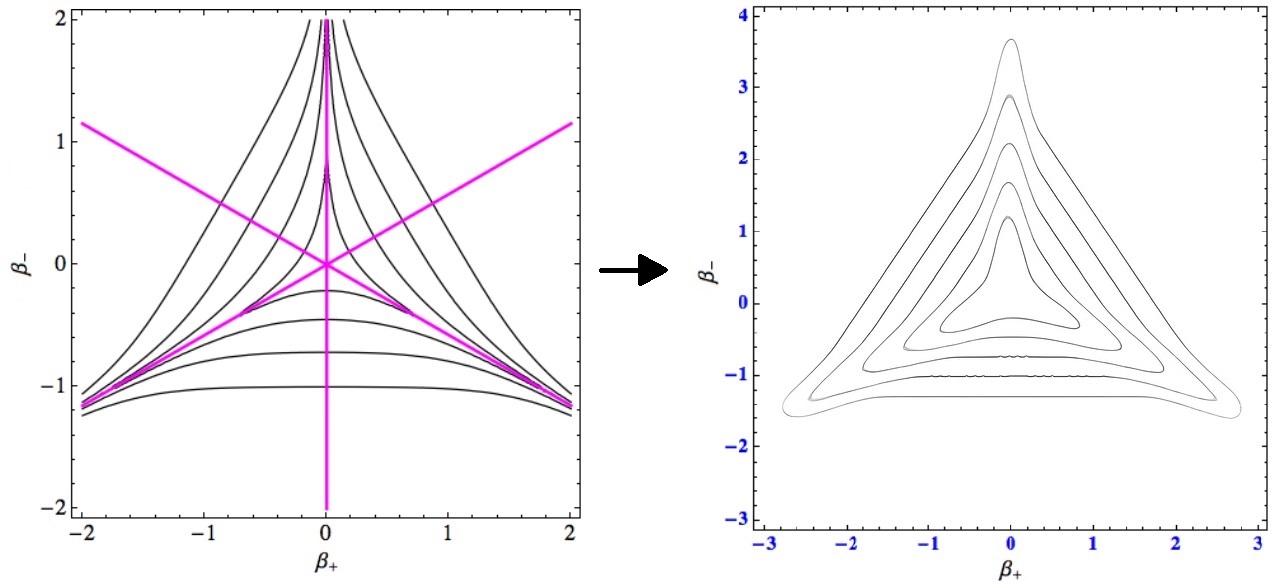}
\caption{\centering\small \textit{Left}: classical anisotropic potential (\ref{b9pot}). \textit{Right}: semi-classical anisotropic potential (\ref{semipot})} 
\label{potentials}
\end{figure}

\subsection{Semi-classical portrait of the isotropy}
In analogy to the Weyl-Heisenberg group for the full plane, we adopt the so-called covariant affine group quantization (see for instance \cite{jpM2016}) of functions defined on the half-plane $\R_+^\ast\times \R$ and apply it to the isotropic variables $(q,p)$. In what follows we make use of the affine coherent states obtained from two distinct fiducial vectors introduced in previous papers \cite{berczugamal20}. One family of the affine coherent states for quantization (obtained from fiducial vectors labelled by $\mu$) and the other one for semi-classical portrait (obtained from fiducial vectors labelled by $\nu$). We find the following lower symbols:
\begin{align}\begin{split}
\reallywidecheck{(p^2)}&=p^2+\frac{K(\mu,\nu)}{q^2},~~K(\mu,\nu)=\left(\frac{\mu+\nu}{2}+\frac{1}{4}\right)e^{\frac{3}{2\mu}},\\
\reallywidecheck{(q^{\alpha})}&=Q_{\alpha}(\mu,\nu)q^{\alpha},~~Q_{\alpha}(\mu,\nu)=e^{\frac{\alpha(\alpha-1)}{4}(\frac{1}{\mu}+\frac{1}{\nu})}.
\end{split}
\end{align}

\subsection{Semi-classical portrait of the total constraint}
The semi-classical portrait of the Hamiltonian constraint \eqref{con} reads as
\begin{align}\label{scconstr}\begin{split}
\reallywidecheck{\mathrm{C}}&=\frac{9}{4}\left(p^2+\frac{K(\mu,\nu)}{q^2}\right)-Q_{-2}(\mu,\nu)\frac{\mathbf{p}^2+\sum_{\pm}\frac{8}{\sigma_{\pm}^2}}{q^2}-36Q_{\frac{2}{3}}(\mu,\nu)q^{\frac{2}{3}}[\reallywidecheck{V}(\bsb)-1]\, .\end{split}
\end{align}
For convenience, we introduce $K_{eff}(\mu,\nu,\sigma_{\pm}):=K(\mu,\nu)-\frac{32}{9}\sum_{\pm}\frac{Q_{-2}(\mu,\nu)}{\sigma_{\pm}^2}>0.$
We derive from the semi-classical Hamiltonian constraint \eqref{scconstr} the following Hamilton equations:
\begin{align}
\label{Ham1}
\dot{q}&=\frac{9}{2}p\,, ~~~~~~~~~~~~\dot{p}=\frac{9}{2}\frac{K_{eff}}{q^3}-2Q_{-2}\frac{\mathbf{p}^2}{q^3}+24Q_{\frac{2}{3}}q^{-\frac{1}{3}}[\reallywidecheck{V}(\bsb)-1]\,,\\
\label{Ham3}\dot{\beta}_{\pm}&=-2Q_{-2}\frac{p_{\pm}}{q^2}\,,~~~~~~~~\dot{p}_{\pm}=36Q_{\frac{2}{3}}q^{\frac{2}{3}}\partial_{\pm}\reallywidecheck{V}(\bsb)\,,
\end{align}
We have thus obtained the semi-classical dynamical system in the full phase space $\R^\ast_+\times \R\times\R^4$ to be now carefully examined. It involves six positive otherwise arbitrary quantization parameters: $\mu, \nu, \sigma_{\pm},\omega_{\pm}$ (degree of confidence...) and produces dynamical trajectories as a function of five initial conditions.

It is straightforward to find the semi-classical versions of the dynamically most relevant geometric quantities in terms of phase space variables:
\begin{align}\label{geoquant}\begin{split}
\check{R}_{iso}=\frac{3Q_{\frac{2}{3}}}{2q^{\frac{4}{3}}},~~\check{R}_{ani}=-\frac{3Q_{\frac{2}{3}}\check{V}(\bsb)}{2q^{\frac{4}{3}}},~~
\check{\sigma}^2=\frac{Q_{-2}\mathbf{p}^2}{48q^4},~~\check{R}_Q=\frac{3K_{eff}}{32 q^4},\end{split}
\end{align}
where $H$, $R_{iso}$, $R_{ani}$ and $\sigma^2$ are respectively the Hubble rate, the isotropic intrinsic curvature, the anisotropic intrinsic curvature and the shear (squared). The semi-classical version of the generalized Friedmann equation reads:
\begin{align}\label{semFried}
H^2=\frac{1}{6}\rho_r-\frac{1}{6}\check{R}_{iso}+\frac{1}{3}\check{\sigma}^2-\frac{1}{6}\check{R}_{ani}-\frac{1}{6}\check{R}_Q,
\end{align}
where $\rho_r=\frac{M_r}{q^{8/3}}$ is the energy density of radiation added to the model.

We interpret the difference between the present semi-classical and the original classical expressions to be the effect of quantum dispersion imposed on the geometry of the universe. The largest discrepancy between the classical and the semi-classical model is given by the repulsive potential $\frac{K_{eff}}{q^2}$ (or, equivalently, the quantum curvature $\check{R}_Q$). Another strong quantum feature is given by modifications to the anisotropy potential $\check{V}(\bsb)$. The notion of a {\it 4-d spacetime} is no longer applicable as the momenta $(p,p_{\pm})$ that define the embedding of the the intrinsic geometry into a spacetime do not commute with the three-geometry variables $(q,\beta_{\pm})$. The {\it geometrical} relation connecting the intrinsic geometry with a higher-dimensional geometry is thus destroyed. Hence, one could argue that quantization of general relativity brings us back to the idea of space as an independent, basic entity. Then, the cosmological singularities become merely an artefact of an incomplete dynamics of three-surfaces, which can now be modified or extended beyond the singular point while avoiding any {\it geometrical} inconsistencies.

\section{Semi-classical dynamics}
\label{semiclassdyn}
\subsection{Isotropic case}
We start by assuming perfectly spherical spatial sections with $\beta_{\pm}=0=p_{\pm}$. Then, only the isotropic part of the constraint is non-trivially vanishing.
A few bouncing trajectories are plotted in Fig. \ref{figure0}. 
We note that in the isotropic case the phase of accelerated expansion is very brief and clearly insufficient from the point of view of the process of structure formation at the substantial range of scales. Indeed, we find that the number of e-folds of inflation is just $N=\ln \frac{a_{end}}{a_{min}}=\ln \sqrt{2}$.

\begin{figure}[h!]\centering
\includegraphics[width=0.45\textwidth]{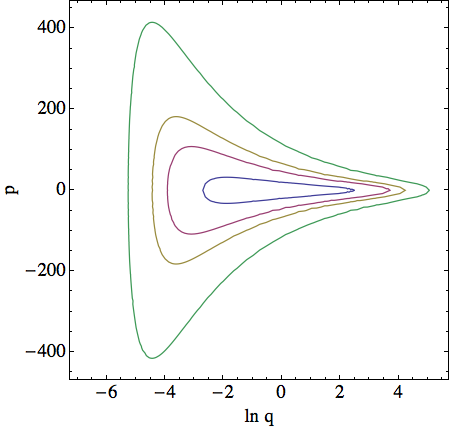}
\caption{\centering\small Isotropic bouncing solutions with varying values of $M_r$ {\small($\sigma_{\pm}=\omega_{\pm}=\mu=\nu=10$)}.}
\label{figure0}
\end{figure}

\subsection{Anisotropic case}
The introduction of anisotropy makes the dynamics of the universe too complex to be solved analytically. Therefore, in the present work \cite{paper} we choose to combine numerical computations with some analytical estimates. In Fig. \ref{fullani} we present an example of numerical results for the generic anisotropic semi-classical mixmaster dynamics.

It is important to keep in mind that this is a chaotic system, very sensitive to initial conditions and the quantization parameters. This means that we can obtain a huge variety of different shapes for the phase-space evolution of the Universe. However, all the solutions share a general timeline: In the typical scenario of the mixmaster universe the quantum potential $K_{eff}/q^2$ diminishes rapidly after the bounce with the anisotropy taking over the dynamics. Eventually, the matter density exceeds the anisotropy, and the standard Friedmann
cosmology begins. One typically observes a few oscillations in the expansion right after the bounce. Contrary to the isotropic case, the bounce is no longer symmetric, leading to the destruction of the cosmic periodicity with each new cosmic cycle being different from the previous one. It can be shown that the bounce must eventually happen for any trajectory. The
observed dynamics points to the possibility for a phase
of sustained accelerated post-bounce expansion lasting
for many e-folds. We shall investigate this issue in the
following section. 

\begin{figure*}[h!]
{\centering
\begin{tabular}{cc}
\includegraphics[width=0.4\textwidth]{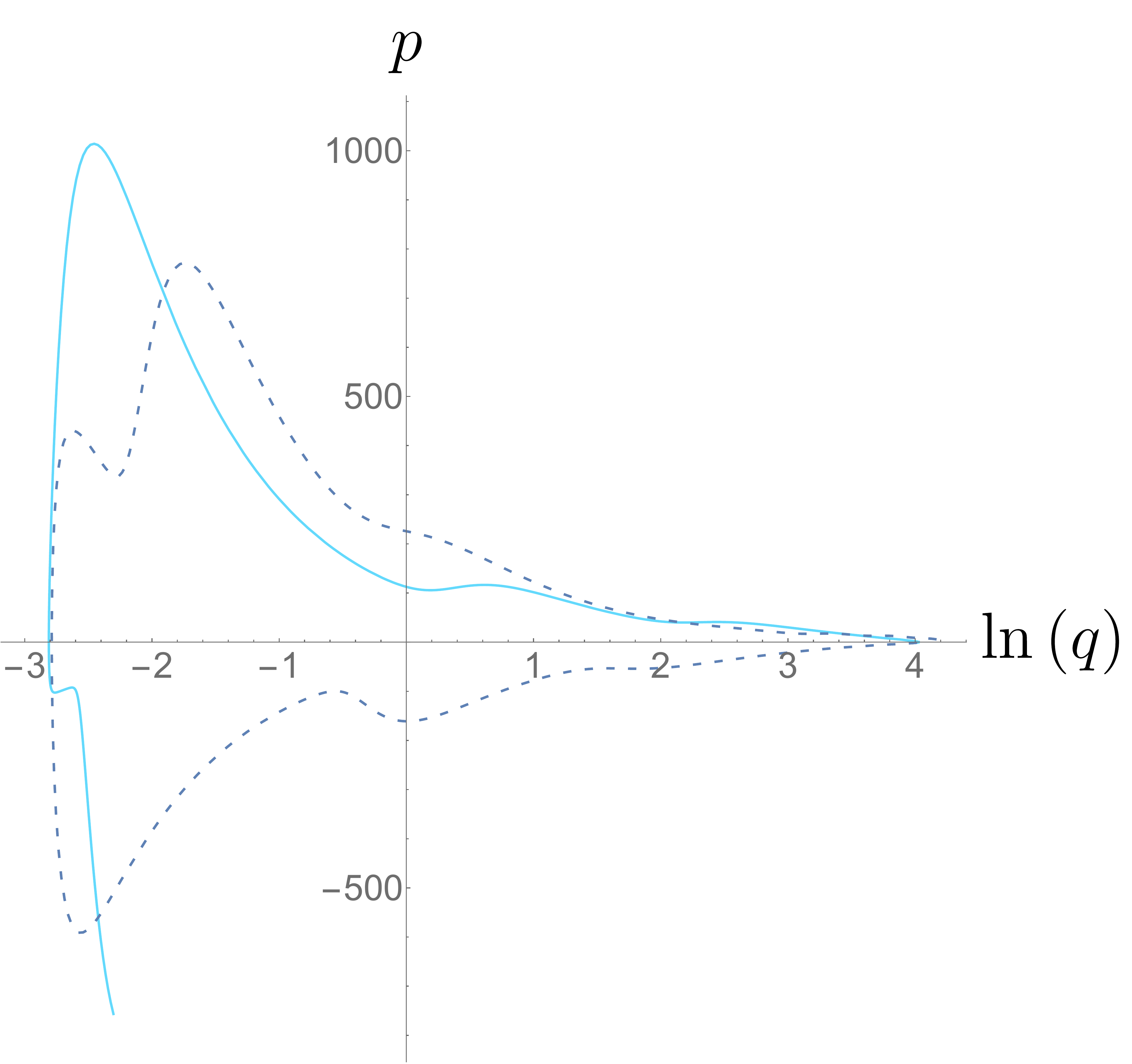}\hspace{1.5cm}
\includegraphics[width=0.4\textwidth]{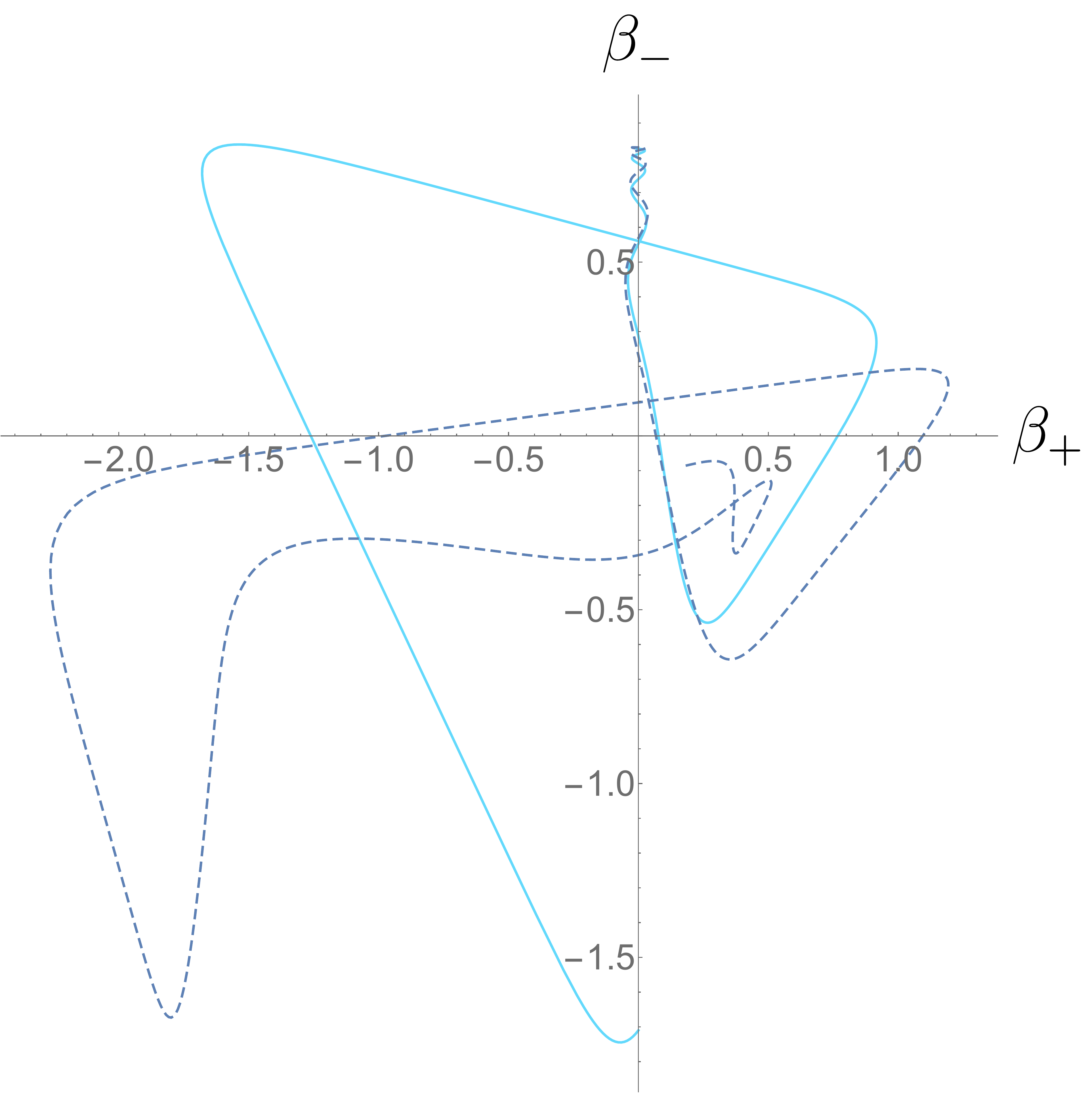}
\end{tabular}
\caption{\centering$\small \text{Initial data:} ~\beta_+(0)=0, ~\beta_-(0)=-1.71, ~p_+(0)=-80, ~p_-(0)=80, ~q=0.1, ~p=-756.41,$ $~M_r=10^2, ~\sigma^2_{\pm}=100, ~\omega_{\pm}=10^5, ~\mu=\nu=10^4.$ 1$^\text{st}$ cosmic cycle (solid line) and 2$^\text{nd}$ cycle (dashed).} 
\label{fullani}
}
\end{figure*}

\section{Accelerated expansion}
\label{accelerated}

The condition that defines if accelerated expansion takes place or not is: 
\begin{align}\label{infl1}
\ddot{a}>0, ~~~~\text{or} ~~~~\mathcal{H},_{\eta}>0,
\end{align}
where $\mathcal{H}=\frac{\acute{a}}{a}=\dot{a}$ is the conformal Hubble parameter. When the conformal Hubble horizon $\mathcal{H}^{-1}$ is shrinking, perturbation modes of fixed co-moving wavelengths $k^{-1}$ leave the horizon and become amplified. It is often assumed that the span of wavelengths that exit the horizon during the inflationary phase is such that $\frac{k_{fin}}{k_{ini}}\gtrsim 10^{8}$.
If there exists an inflationary dynamics in the semi-classical mixmaster model it must be driven by either the quantum curvature $\frac{1}{6}\check{R}_Q$ or the anisotropy energy $\frac{1}{3}\check{\sigma}^2-\frac{1}{6}\check{R}_{ani}$, or a combination of both.

 Let us now assume that at each moment of time the terms of generalized Friedmann equation are well-approximated by power functions, $\frac{1}{6}\check{\rho}_{ani}=\frac{\lambda_{ani}}{a^{n_{ani}+2}}$ and $\frac{1}{6}\check{R}_Q=\frac{\lambda_{Q}}{a^{6}}$, that is, they are linear with respect to the number of e-folds. In order for the accelerated expansion to occur Eq. \eqref{infl1} must hold, that is,
$0<\lambda_{Q}-\frac{n_{ani}\lambda_{ani}}{4a^{n_{ani}-4}},$ which must also be consistent with the Friedmann equation, $0<\frac{\lambda_{ani}}{a^{n_{ani}-4}}-\lambda_{Q}.$ In the case of both anisotropy and quantum curvature contributions being important, the above conditions can be easily combined (see \cite{paper} for more details). By doing so, it is straightforward to see that $0<n_{ani}<4$. The above conditions must hold for around $\Delta N=20$ e-folds (considered a sufficient amount in inflationary models). We have $e^{\Delta N(4-n_{ani})}=\frac{4}{n_{ani}}$, and hence $n_{ani}$ has to be very small. Note that for $n_{ani}\approx 4$ we obtain the minimal number of e-folds $\Delta N=0.25$.  

By setting $\mathcal{N}:=(-36Q_{\nicefrac{2}{3}}(\mu,\nu)q^{\frac{2}{3}})^{-1}$, the semi-classical Hamiltonian \eqref{scconstr} acquires the standard form for a free particle inside our potential,
where the mass is $m(q)=\nicefrac{18Q_{\nicefrac{2}{3}}(\mu,\nu)}{Q_{-2}(\mu,\nu)}q^{\nicefrac{8}{3}}$. Hence, the equation of motion for $\ddot{\beta}_{\pm}$ just depends on a friction term proportional to the derivative of the potential, and a second one to the anisotropic momentum.
The kinetic energy scales as $a^{-4}$, whereas the potential energy is independent of the scale factor. Hence, we obtain again $4>n_{ani}>0$. In order to reproduce the inflationary dynamics, we must have $n_{ani}=4e^{-4\Delta N}$, which is very small, for $\Delta N\approx 20$ e-folds. This demands the dynamics to be
dominated by the anisotropy potential with a negligible kinetic energy $\dot{\beta}_{\pm}\approx 0$. In other words, the relative change of the potential during that number of e-folds must be very small. 

\subsection{Can the interplay between anisotropy and semiclassical bounce \\generate a sufficiently long inflationary phase?}

In order to answer this question, firstly, we will make use of numerical simulations to study the most convenient scenario in which the above conditions occur and might be preserved for some enough amount of time. For such a purpose, it is important to know the shape of the potential. $\reallywidecheck{V}(\bsb)$ As we can observe in Fig. \ref{potentials}-\textit{right}, the potential has three ``canyons" in each of the vertices of its \textit{triangular} shape. The length of these canyons can be modulated by the quantization parameters $\omega_{\pm}$. By increasing this parameter, we make the canyons longer, at the same time that the slope of the steep walls of the potential is reduced.

\begin{figure*}[h!]
{\centering
\begin{tabular}{cc}
\includegraphics[width=0.40\textwidth]{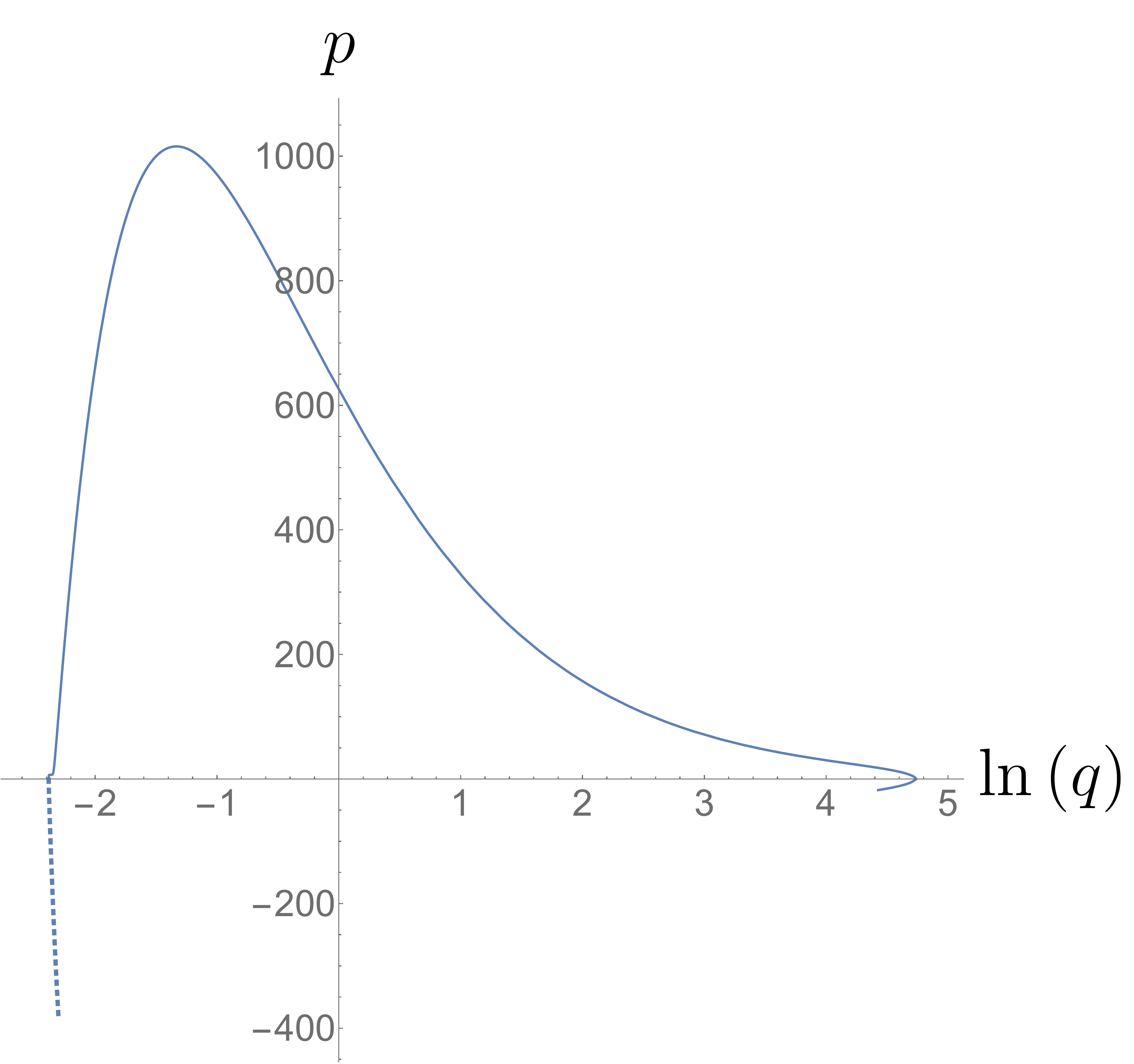}\hspace{1.5cm}
\includegraphics[width=0.40\textwidth]{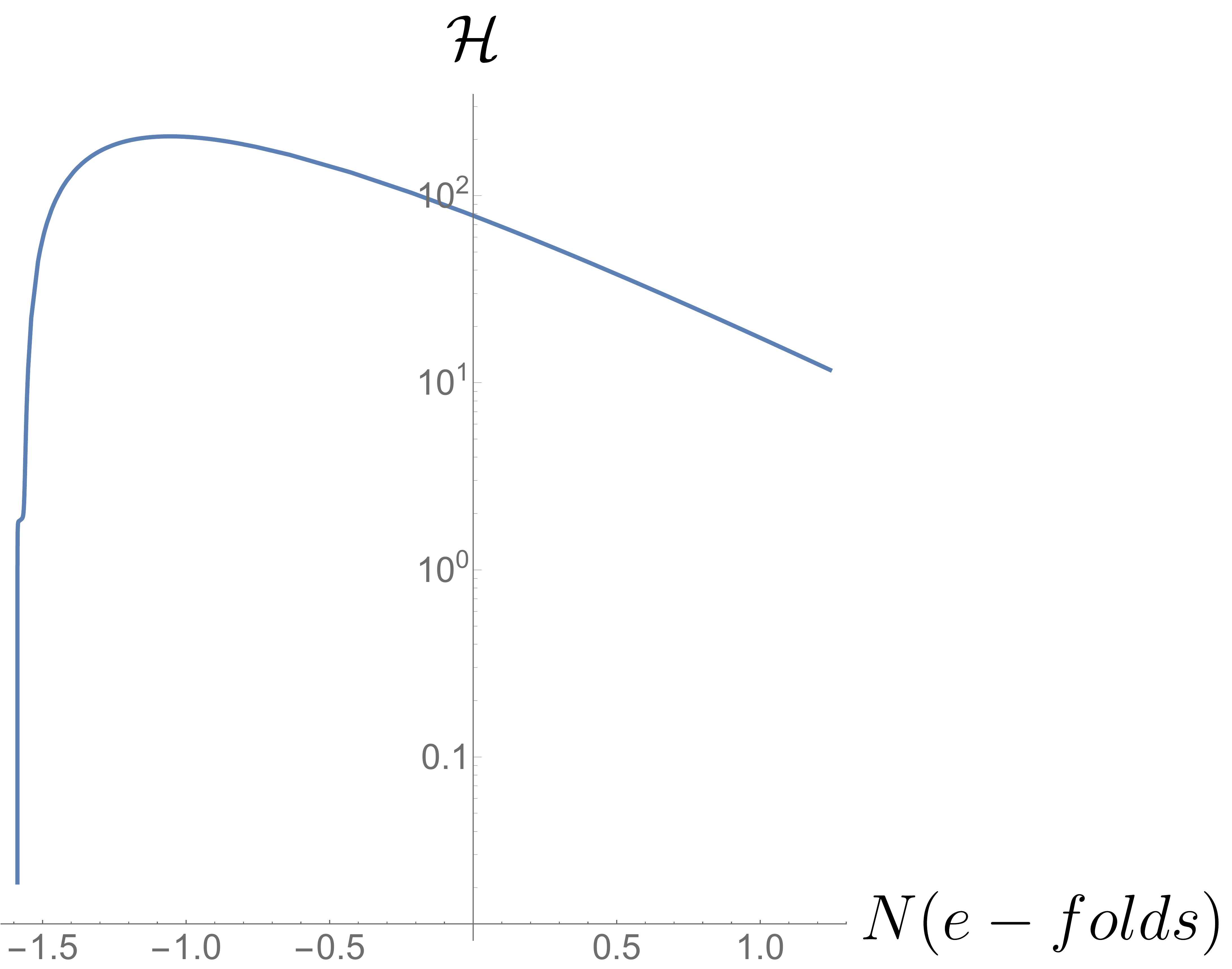}\\
\includegraphics[width=0.40\textwidth]{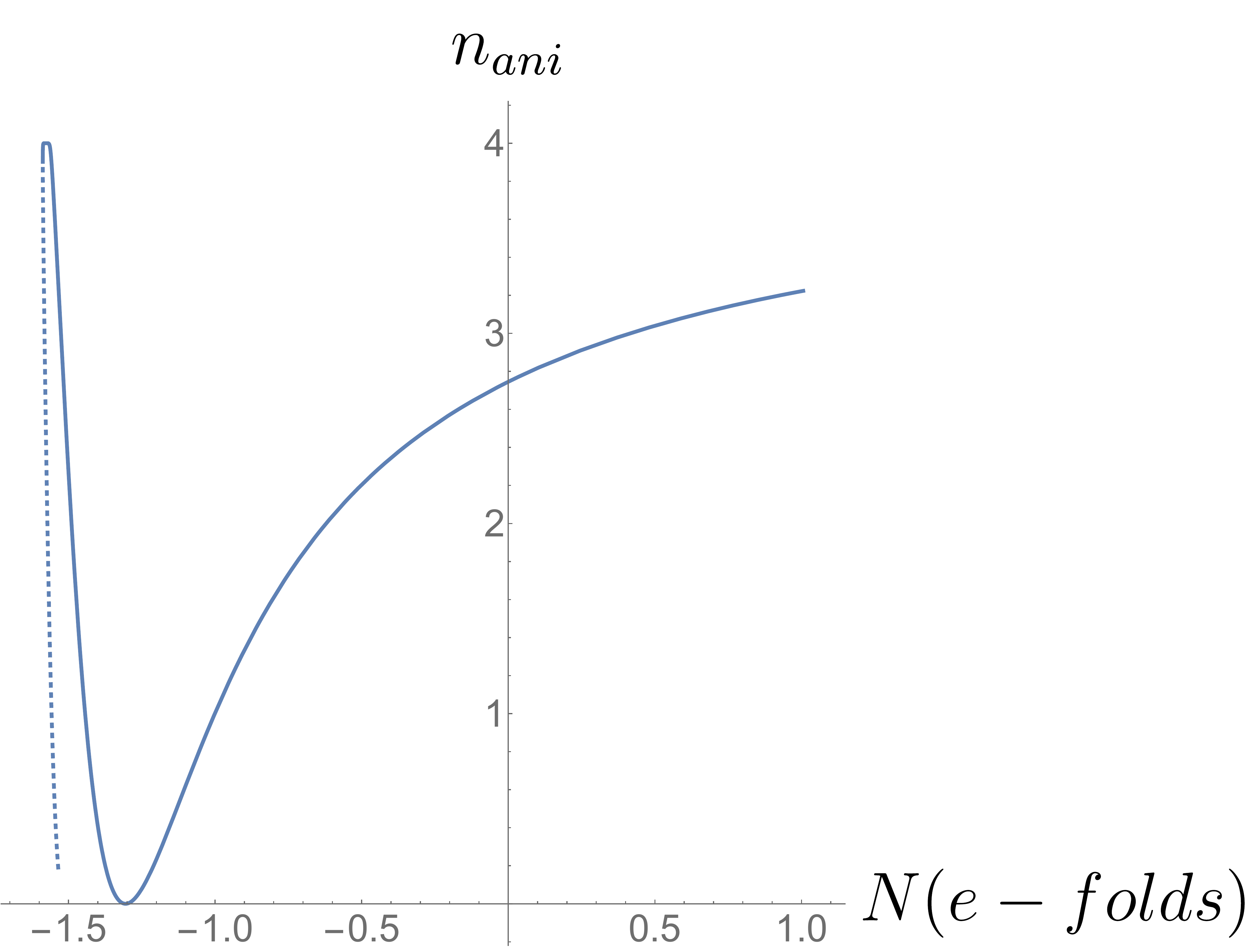}\hspace{1.5cm}
\includegraphics[width=0.40\textwidth]{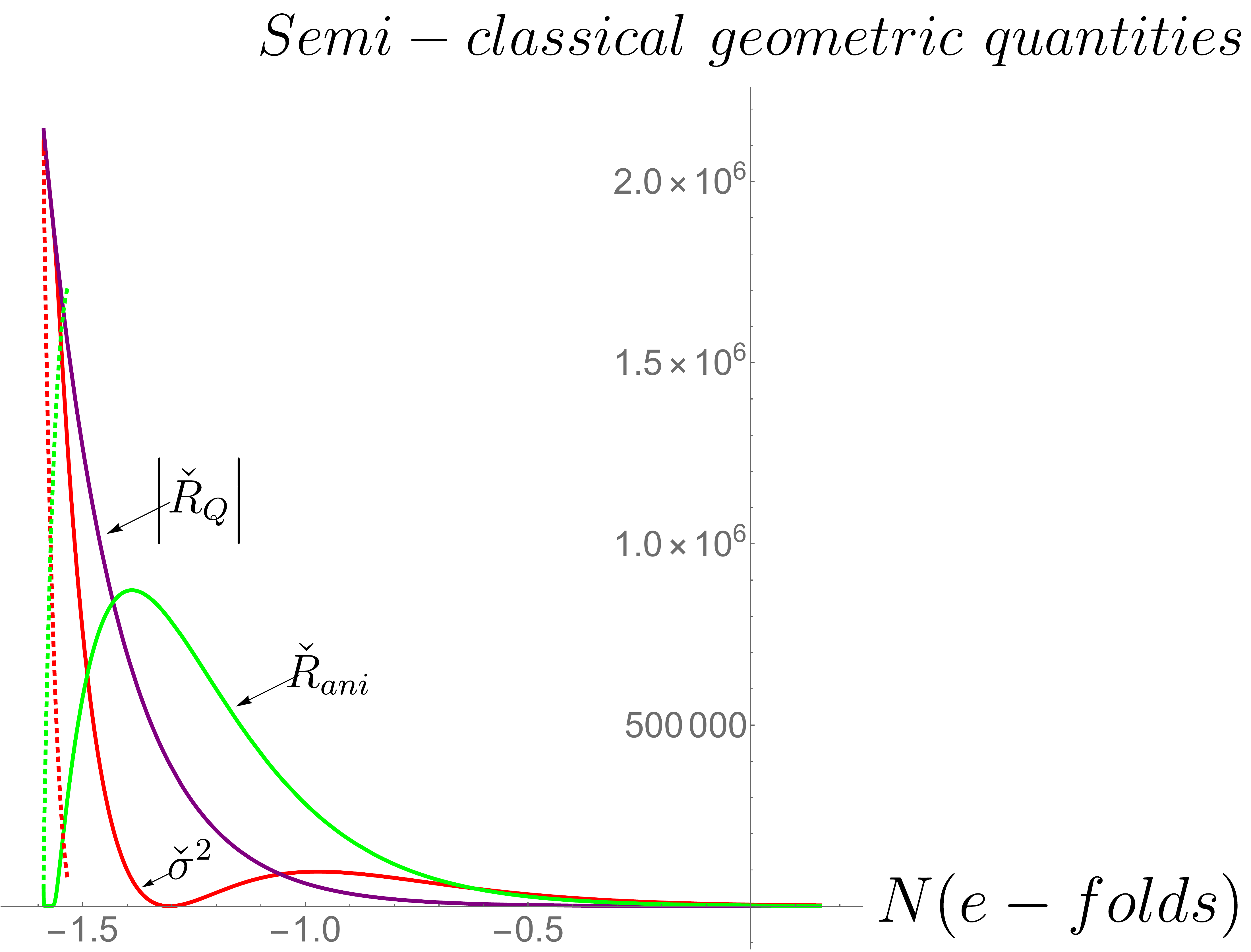}
\end{tabular}
\caption{{\centering \small \textit{Top left}: isotropic phase space expansion. \textit{Top right}: $\cal{H}$ evolution in terms of e-folds of expansion. \textit{Bottom left}: $n_{ani}$ evolution during expansion. \textit{Bottom right}: $\left|\check{R}_Q\right|$, $\check{R}_{ani}$ and $\check{\sigma}^2$ evolution during expansion.}\\ $\small \text{Initial data:} ~\beta_+(0)=0, ~\beta_-(0)=-1.71, ~p_+(0)=0, ~p_-(0)=35, ~q=0.1, ~p=-377.31, ~M_r=10^2$, $~\sigma^2_{\pm}=100, ~\omega_{\pm}=56.23, ~\mu=\nu=10^4.$ \small Expansion phase (solid line) and end of contraction (dashed).} 
\label{acceleration}
}
\end{figure*}

Therefore, the most convenient scenario would consist in making the particle inside the potential to be rolling up (the deepest possible) through one of these canyons, for instance the  one at the top central part of Fig. \ref{potentials}-\textit{right}. Then, we make this situation coincide with the moment when the total anisotropy contribution takes over the quantum potential one for driving the dynamics, such that at that moment the momentum decreases slowly to zero $\dot{\beta}_{\pm}\approx0$ (at the top of the canyon) and the potential is sufficiently high, hence the latter will dominate the dynamics. This is due to the fact that these canyons are the parts of the potential where the slope is not very big, and the relative change of the potential can be maintained small for longer time while particle is rolling up and down the canyon, and at the same time the total value of the potential is big enough. In that way the $n_{ani}\approx0$ situation can be preserved for longer time.

In Fig. \ref{acceleration} we show the results for simulations within such scenario. As we observe in the evolution of $\cal{H}$, there exists an extra boost to the post bounce accelerated expansion just after the quantum potential stops dominating, allowing an increased number of modes to leave the horizon. However, the condition $\cal{\acute{H}}>$ $0$ cannot be extended for an arbitrarily long time, being $N\approx0.54$ the maximum amount of e-folds we could obtain in our simulations.

It can be shown \cite{paper} that requiring the anisotropic potential to lead the accelerated expansion implies that the quantity $\nicefrac{|\reallywidecheck{V}_{,\pm}|}{|\reallywidecheck{V}|}$ must be really small. However, it is also easy to see from the shape of the potential that $2<\nicefrac{|\reallywidecheck{V}_{,\pm}|}{|\reallywidecheck{V}|}<8$ except close to the point of origin $\bsb=0$, where the potential $\reallywidecheck{V}$ has the minimum. We then conclude that the semi-classical potential cannot satisfy the above requirements and thereby excluding a sufficiently long inflationary phase from this model.

\vspace{-0.75ex}
\section{Conclusions}
\label{disc}
In this work we investigated whether a quantum anisotropic universe can mimic the dynamical behaviour of an inflationary solution, spontaneously inducing a sufficiently long primordial phase of accelerated expansion. We first derived a quite generic quantum model of mixmaster universe via integral covariant quantization method and coherent states. Then using its equations of motion we found the reasons for why anisotropic universe, neither classical nor quantum, can not induce a sustained inflationary phase in the early universe. 

Nevertheless, the analysis that we performed in this work was semi-classical and perhaps going to a fully quantum description could change the character of the bouncing solutions. In addition, if we included the backreaction from quantum perturbations, the
anisotropy potential could perhaps acquire large corrections allowing for sustained inflationary phase. We could also propose another cosmological scenario in which anisotropy plays a key role in the generation of structure from primordial perturbations, with a
bouncing cosmology in which the generation starts in the contracting phase and then is smoothly transferred through a bounce to the expanding phase. Such cosmology was already proposed in \cite{mipaper} for an FLRW background. The addition of anisotropy could modified the blue-tilted power spectrum produced by the isotropic model \cite{mipaper2}. The results showed in this contribution will be presented in the forthcoming paper \cite{paper}, including a more exhaustive analysis of the model and rigorous discussion.

\vspace{-2ex}
\section*{Acknowledgments}
The author acknowledges the support of the National Science Centre (NCN, Poland) under the research grant 2018/30/E/ST2/00370.


\end{document}